\documentstyle[aaspp4]{article}
\begin{document}
\singlespace

\title{A possible route to spontaneous reduction of the heat conductivity 
by a temperature gradient driven instability in electron-ion plasmas}
\author{Makoto Hattori}
\affil{Astronomical Institute, T\^{o}hoku University, Aoba,
Sendai 980-8578, Japan; hattori@astr.tohoku.ac.jp}
\authoraddr{Astronomical Institute, T\^{o}hoku University, Aoba,
Sendai 980-8578, Japan}
\and
\author{Keiichi Umetsu}
\affil{Astronomical Institute, T\^{o}hoku University, Aoba,
Sendai 980-8578, Japan; keiichi@astr.tohoku.ac.jp}
\received{July, 1999}
\accepted{November, 1999 in ApJ}
\journalid{}{}

\begin{abstract}
We have shown that there exists low-frequency growing modes
driven by a global temperature gradient 
in electron and ion plasmas, by linear perturbation analysis 
within the frame work of plasma Kinetic theory.
The driving force of the instability is the local 
deviation of the distribution function from the Maxwell-Boltzmann 
due to global temperature gradient. 
Application to the intracluster medium shows that 
scattering of the particles due to waves excited by the instability 
is possible to reduce
mean free paths of electron and ion down to five to seven order of magnitude
than the mean free paths due to Coulomb collisions.
This may provide a hint to explain 
why hot and cool gas can co-exist in the intracluster medium
in spite of the very short evaporation time scale due to  
thermal conduction if the conductivity is 
the classical Spitzer value.   
Our results suggest that the realization of the global thermal
equilibrium is postponed 
by the local instability which is induced for quicker 
realization of local thermal equilibrium state in plasmas. 
The instability provides a new possibility to create and grow 
cosmic magnetic fields without any seed magnetic field.
\end{abstract}


\section{Introduction}

One of the most important results obtained 
by Japanese/US X-ray satellite ASCA 
(Tanaka, Inoue, \&  Holt 1994)
is the first spectro-scopical confirmation  of  existence of  
at least two different temperature X-ray emitting hot gas in the cluster
central region (Fukazawa et al. 1994; 
Ikebe 1995; Ikebe et al. 1999);
the cooler is about 1keV and the hotter is about 2keV--10keV.
Although the interpretation of the result is still controversial (Fabian 1994; 
Ikebe et al. 1999), there is a 
common feature among the different interpretations, 
that is, a cool region with temperature $T_1$ is
surrounded by a hot region with temperature $T_2$ 
maintaining a pressure equilibrium (see \S 6.2). 
Since the total thermal energy in the hot region is much 
higher than that in the cool region, the hot region can be 
treated as a heat reservoir.
Suppose that the interface separating two plasma regions
with different temperatures
has a thickness of $L$ and temperature variation in the interface region 
is almost linear in spatial scale.
Assume that the electron mean free path $\lambda_e$ is smaller than $L$.   
The time scale $\tau_{\rm ev}$ of the evaporation 
due to heat conduction  can be roughly estimated by a time scale 
for electrons in the hotter region to diffuse across the interface,
and is given as 
\begin{equation}
\label{eq:diftim}
\tau_{\rm ev}\sim (\lambda_e/v_{{\rm
th},e})\times (L/\lambda_e)^2
=\epsilon^{-1} (L/v_{{\rm th},e})\sim 10^5(L/{\rm 4kpc})
(k_{\rm B} T_e/4{\rm keV})^{-0.5}(\epsilon/1.0)^{-1}{\rm yr},
\end{equation} 
where $v_{{\rm th},e}\equiv \sqrt{2 k_{\rm B} T_e/m_e}$ is the 
electron thermal velocity, $T_e$ is the electron temperature, 
and $m_e$ is the electron mass; 
$\epsilon$ is defined by $\epsilon \equiv \lambda_e/L$.
In the cluster central region, the electron mean free path due to
Coulomb collision is obtained as $\lambda_{\rm C}\sim 0.5(k_{\rm B} T_e/4 
{\rm keV})^2 (n_e/10^{-2}{\rm cm^{-3}})^{-1}{\rm kpc}$ (Sarazin 1988). 
Therefore, the time scale of the evaporation is about $10^6$ yr 
and is much shorter than the age of cluster of galaxies,
which is about the Hubble time.
For maintaining the interface at least as long as the age of the cluster, 
the reduction of the electron mean free path 
in four order of magnitude from the Coulomb mean free path
to realize $\epsilon$ of $\sim 10^{-5}$ is required. 
Soker \& Sarazin (1990) and Makishima (1997) proposed that
the cool and hot phases are thermally insulated from each other by magnetic 
fields.  However, their models involve another problem 
how such special magnetic field structures  are obtained  
in the cluster central region. 
In this paper, we show that there is an alternative possibility to reduce
electron mean free path in many order of magnitude 
without any special assumption on magnetic field structure.

Ramani \& Laval (1978) found a new plasma instability 
which may relate to the reduction of the electron mean free path,
henceforce the reduction of the heat conduction. 
They showed that the temperature gradient leads to an anisotropic 
electron velocity distribution function and that the anisotropy 
of the velocity distribution function drives the 
instability like the Weibel instability (Weibel 1959; Fried 1959; 
Melrose 1986).
They proposed that the chaotic magnetic field and electric field,
produced by the plasma wave induced by the instability, 
scatter electrons, so that the electron mean free path
is reduced in many order of magnitude. 
However, the instability found by them cannot be applied 
to reduce the electron mean free path in many order of magnitude
because of the following reasons. 
Since they assumed that only electrons respond
to the mode and ions were treated as fixed back ground particles, 
the phase velocity of the wave 
must be faster than the ion sound velocity $v_{{\rm th},i}=
\sqrt{2 k_{\rm B} T_i/m_i}$, 
%and/or the ion plasma frequency
%$\omega_{\rm pi}=\sqrt{4\pi n_i e^2/m_i}$, 
where $T_i$, $n_i$, 
and $m_i$ are the ion temperature, ion number density, 
and ion mass, respectively.  
In the astrophysical plasma, 
the ion mass can be safely replaced by the proton mass.    
The unstable mode found by Ramani \& Laval (1978) 
has non zero real part of the wave frequency with a phase velocity  
of $\sim \epsilon v_{{\rm th},e}$.
Therefore, the application limit of their analysis
sets a relatively high lower limit on $\epsilon$
as $\epsilon > \sqrt{m_e/m_i}\sim 0.025$ when $T_i=T_e$
(hereafter we refer to this lower limit as 
the wall of the square root of the mass ratio). 
To explain the two-phase nature of the hot gas in cluster central region,
the instability found by them is practically not
useful and the mechanism which can break  
the wall of the square root of the mass ratio is required.

The first application of the Ramani \& Laval (1978) type instability to the
astrophysical plasma was made by Levinson \& Eichler (1992).
They extend the Ramani \& Laval (1978)'s analysis to include 
the non-zero background magnetic field and 
found the low-frequency unstable mode, similar to the Ramani \& Laval (1978)
type modes.
Pistnner, Levinson \& Eichler (1996) first applied 
the Levinson \& Eichler (1992)'s results to the cluster cooling flow.
However, they applied their results to reduce electron mean free
path down to $10^{-3}$ and far less than the lower limit, 
bounded by the application limit 
of the wall of the square root of the mass ratio. 
Further they neglected the cluster gravitational force
in their Boltzmann equation in spite of taking into account  
a pressure gradient with a scale height comparable to 
the cluster gravitational scale height. 
As a result, they  implicitly assumed the existence of the 
steady electric field for the electric force acting
on an electron to balance with the pressure gradient force acting on the electron. 
Since the pressure gradient force acting on an ion
balances with the cluster gravitational force acting on the ion,
the electric force 
acting on an ion is as strong as the cluster
gravitational force acting on the ion. 
The existence of such a strong electric field in cluster of galaxies
is unlikely.
Further, the model adopted as the temperature distribution in 
the cluster central region obtained by White \& Sarazin (1987)
only describes the temperature distribution of the hottest phase gas.
Therefore, the model adopted by them is inappropriate to 
answer 
why the multi-phase gas clumps can 
co-exist in the cluster of galaxies.

In this paper, we extend the Ramani \& Laval (1978)'s analysis
including the response of the ion to examine whether we can break the
wall of the square root of the mass ratio. 
We limit our attention to the pressure equilibrium plasma;
in other words, the scale of the interface is much smaller than the
pressure scale height. 
Then, the gravitational force can be neglected. 
The plan of the paper is as follows. 
The readers who are not interested in the details of the
derivation of the dispersion relation can skip \S\S 3 and 4. 
In \S 2, we set the model, and summarize the basic assumptions 
and the application limit coming from the assumptions.  
In \S 3, we deduce the solution of the Boltzmann equation 
under the situations set in \S 2 using the Chapman-Enskog
expansion of the electron and ion velocity distribution functions. 
We deduce the dispersion relations under low-frequency 
conditions in \S 4.
In \S 5, we summarizes the characteristics of the instability and modes
found in \S 4. 
The application of the results to the hot plasma in the central region of
cluster of galaxies
is discussed in \S 6. 
Finally, \S 7 summarizes another important implications from our results
and remained problems.

\section{The model and application limit}	

Consider the situation described in Fig. 1. 
Two plasma regions with different temperatures  
contact maintaining pressure equilibrium. The electron temperature
in the transition region with a depth of $L$
varies almost
linearly from $T_1$ to $T_2$ ($T_1<T_2$) along the $x$-direction:
$\nabla T(x)/T(x)=\delta_T L^{-1}$ where $\delta_T\equiv (T_2-T_1)/T$. 
In clusters of galaxies, the pressure scale height
is about the core radius of the cluster mass distribution,
say $\sim 50$kpc (e.g., Hattori, Kneib, \& Makino 1999).
Therefore, if the size of the interface is smaller than $\sim 10$kpc,
the pressure equilibrium
%isobaric 
condition for the interface region can be safely applied.
The ion temperature is assumed 
to be the same as electron temperature,$T_i=T_e$. 
%The mean free paths of electron, $\lambda_{{\rm m},e}$,  and ion, 
%$\lambda_{{\rm m},i}$, 
The mean free paths $\lambda_e$ and $\lambda_i$ of electron and ion,
respectively, 
are assumed to be defined by the scattering due to plasma waves.
If the scattering of the particle due to electric field $E$ 
of the wave is the dominant source, the mean free path of  
each particle could be described as follows.
Assume that the electric field associated with plasma wave 
is coherent in the scale of order of the wave length
$\lambda$ and is chaotic over the scale. 
Then the momentum change of the particles can be described by the 
random walk manner in the momentum space.  
The mean free path is defined as the length scale 
that the particles gain or loss its momentum 
comparable to their original momentum during they move 
the scale.  The momentum change of electron during one step is 
$m_e \Delta v_e \sim e E \lambda/v_{{\rm th},e}$.
The number of steps required for electron to gain or loss
their momentum comparable to its original momentum is
$[m_e v_{{\rm th},e}/(e E \lambda/v_{{\rm th},e})]^2
\sim (2 k_{\rm B} T_e/ e E \lambda)^2$.
Then we obtain 
\begin{equation}
 \label{eq:lem}
\lambda_e \sim \left({2 k_{\rm B} T_e\over e E\lambda}\right)^2\lambda
\sim \left({2 k_{\rm B} T_e\over e E}\right)^2
{1\over \lambda}. 
\end{equation}
Similarly, the mean free path for ion can be obtained  as
\begin{equation}
\label{eq:lim}
\lambda_i 
\sim \left({2 k_{\rm B} T_i\over e E\lambda}\right)^2\lambda
\sim \left({2 k_{\rm B} T_i\over e E}\right)^2
{1\over\lambda}. 
\end{equation}
Since we are now assuming that electron and ion have the same temperature,
the electron and ion mean free paths due to the scattering by the 
wave electric field are identical.
When the chaotic magnetic field associated with the plasma wave
is the dominant source of the particle scattering,
the particle mean free path would be defined in the following way.
The magnetic moment of the particles, $\mu=(1/2 m_e v_{\perp}^2)/B$, 
is a conserved variable in
a magnetized plasma, 
where $v_{\perp}$ is the particle velocity perpendicular to 
the magnetic field, written as 
$m_e v_{\perp}^2 \sim 2 k_{\rm B} T_e$, 
just after the perturbed magnetic field is induced.  
As the magnetic-field strength gets stronger, 
$v_{\perp}$ increases to conserve $\mu$.  
The maximum allowable value 
of $v_{\perp}$ is $\sqrt{3 k_{\rm B} T_e/m_e}$, since the kinetic energy
of the particle must be conserved. 
Then it follows that 
an electron at the place with a magnetic field strength of $B_{\rm 1}$
is scattered back at the place with a magnetic 
field strength of $B_{\rm cr} \sim 3/2 B_{\rm 1}$ 
owing to the mirror effect.   
The mean free path is then defined as the length scale along the magnetic 
field from the place where the magnetic field 
strength is $B_1$ to the place where the magnetic field strength 
gets $B_{\rm cr}$. 
Since the critical value $B_{\rm cr}$ of the magnetic field
strength, in other words the place of the bouncing, 
does not depend on the particle species and only depends on 
their initial position and the magnetic-field configuration.
Therefore, the mean free paths of ion and electron must be also
the same in this case.  
These arguments show that the mean free paths of electron
and ion due to scattering by plasma waves may 
have the same order of magnitude.  
Therefore, we assume that electrons and ions 
have a similar value of $\epsilon$.

%As shown by Ramani and Laval (1978), 
%the distribution function is an odd function of the velocities, 
%the real part of the frequency does not vanish and satisfies 
%\begin{equation}
%\label{eq:cond1}
%|\omega_{\rm r}| \sim \epsilon k v_{{\rm th},e} \ll k v_{{\rm th},e}.
%\end{equation}
%Since Ramani and Laval (1978) neglected the motion of ion, 
%the phase velocity of the wave cannot be smaller than the 
%ion sound velocity and this condition leads 
%\begin{equation}
%\label{eq:cond2}
%\epsilon \gg (m_e/m_i)^{1/2},
%\end{equation}
%if $T_e = T_i$.
%To see whether the wall of the square root of the mass ratio can be
%broken if the motion of ion is allowed, we further extend the 
%low frequency condition and try to find the mode  
%in which the phase velocity of
%the wave is much smaller than the ion thermal velocity, $v_{{\rm th},i}=
%\sqrt{2 k_{\rm B} T_i/m_i}$.

%Further, to neglect collisions in the equations of the perturbed variables, 
%a short wavelength condition for the waves such as
In this paper, perturbations with a short-wavelength wave, such that
\begin{equation}
\label{eq:cond3}
\lambda \ll \lambda_e,
\end{equation}
are treated. Therefore, the collision term in the Boltzmann equation
for the perturbed variables can be neglected.

Self-consistency check of the obtained results with all these conditions
constrain the allowable range of $\epsilon$.  
In addition, two further conditions provide important limits 
on $\epsilon$. 
First of all, the electron mean free path due to Coulomb collision 
must be shorter than $L$ at least before the onset of the 
instability, otherwise there is no chance for 
the electron distribution function to have anisotropy 
as described in the next section, and hence no chance to have instability. 
This condition yields an lower limit on $L$ as
\begin{equation}
\label{eq:coulomb}
L> 0.5 (k_{\rm B}T_e/4 {\rm keV})^2(n_e/10^{-2}{\rm cm^{-3}})^{-1}{\rm kpc}.
\end{equation}
Secondly, 
the growth time scale of the unstable mode, $\omega_{\rm i}^{-1}$,
must be shorter than the time scale 
of thermal diffusion when waves do not exist, 
%otherwise there is no chance for the waves to grow
in order for the waves to grow
before the temperature difference will be erased owing to 
quick thermal conduction.
This condition provides
\begin{equation}
\label{eq:cond4a}
\omega_{\rm i}^{-1} < \epsilon^{-1} L/v_{{\rm th},e},
\end{equation}
where only the case that the electron mean free path is shorter than
$L$ is considered.
A further condition has to be required when applied to the clusters, 
that is 
the growth time scale  must be shorter than
the age of the object, conservatively the age of the universe:
\begin{equation}
\label{eq:cond5}
\omega_{\rm i}^{-1}< 1/H_0 =10^{10}h^{-1}{\rm yr},
\end{equation}
where $H_0=100h{\rm km/sec/Mpc}$ is the Hubble constant.

%Another condition is that the growth time scale  must be shorter than
%the age of the object, conservatively the age of the universe such as
%\begin{equation}
%\label{eq:cond5}
%\omega_{\rm i}^{-1}< 1/H_0 =10^{10}h^{-1}{\rm yr},
%\end{equation}
%where $H_0=100h{\rm km/sec/Mpc}$ is the Hubble constant.

\section{Distribution Function}

When $\epsilon <1$, the discussion given in the previous section  
ensures that the  solution of the Boltzmann equation 
of electron and ion can be expanded in powers of $\epsilon$ 
(Chapman \& Cowling 1960)
\begin{eqnarray}
 \label{eq:f}
f_e &=& f_{{\rm m},e}+ f_e^{(1)}+f_e^{(2)}+...,\\
f_i &=& f_{{\rm m},i}+ f_i^{(1)}+f_i^{(2)}+...,
\end{eqnarray}
where $f_{{\rm m},e}$ and $f_{{\rm m},i}$ are Maxwellian distribution
functions for electron and
ion, respectively;
$f_e^{(k)}$ and $f_i^{(k)}$ ($k=1,2,\cdots$) describe the deviation of 
distribution
functions from the Maxwellian in order of $\epsilon^k$ for electron and
ion, respectively. 
Since we assume that $T_i(x)=T_e(x)\equiv T(x)$ and 
the charge neutrality must be maintained, 
the Maxwellian parts for both electron and ion are  written as 
$f_{{\rm m},e}=n_0(x)[\pi v_{{\rm th},e}(x)]^{-3/2} 
\exp\{-[v/v_{{\rm th},e}(x)]^2\}$ and 
$f_{{\rm m},i}=n_0(x)[\pi v_{{\rm th},i}(x)]^{-3/2} 
\exp\{-[v/v_{{\rm th},i}(x)]^2\}$, respectively.
Here $n_0(x)$ is the electron number density, 
$v_{{\rm th},e}(x) \equiv \sqrt{2 k_{\rm B} T(x)/m_e}$, 
and $v_{{\rm th},i}(x)\equiv \sqrt{2 k_{\rm B} T(x)/m_i}$.

The pressure equilibrium assumption gives 
\begin{equation}
\label{eq:peq}
{\nabla n_0\over n_0}=-{\nabla T\over T}=-{1\over L}\delta_T.
\end{equation}
Once this condition is satisfied, the pressure is 
time independent everywhere, even if a secular variation of the temperature 
due to thermal conduction is taken into account.  
Then the time dependence of the density can be related to 
that of the temperature as
\begin{equation}
\label{eq:ntd}
{1\over n_0}{\partial n_0\over \partial t}= 
-{1\over T}{\partial T\over \partial t}.
\end{equation}

The Boltzmann equation under the pressure equilibrium condition 
without background electric and magnetic fields leads to
\begin{equation}
\label{eq:Beq}
{\partial f_e\over \partial t}+v_x {\partial f_e\over \partial x} = 
\left({\partial f_e\over \partial t}\right)_{\rm c},
\end{equation}
where $(\partial f/\partial t)_{\rm c}$ is the collision term
and the equilibrium electric field does not appear because
of the pressure equilibrium condition.
The secular time variation of the distribution function
is ascribed to the secular variation 
of the temperature due to thermal conduction.
Since the cool region with a temperature of $T_1$ is 
considered to be immersed in a heat bath with a temperature of $T_2$,
the temperature of the interface increases monotonically 
owing to the heat conduction.  
Hence, the time evolution of temperature can be described as 
\begin{equation}
\label{eq:tvar}
{\partial T\over \partial t} = 
\epsilon \delta_T{v_{{\rm th},e}\over L}T. 
\end{equation}
Then,
\begin{equation}
\label{eq:dfdt}
\left|{\partial f_e\over \partial t}\right|\sim {1\over T} \left|
{\partial T\over \partial t}\right| f_e
\sim \epsilon \delta_T {v_{{\rm th},e}\over L} f_e.
\end{equation}
The simplest choice of the collision term is the Krook operator, such that
\begin{equation}
\label{eq:Krook}
\left({\partial f_e\over \partial t}\right)_{\rm c}=
-\nu_e (f_e-f_{{\rm m},e}),
\end{equation}
where $\nu_e=v_{{\rm th},e}/\lambda_e$ is the collision frequency
assumed to be constant.
Substituting the collision term by the Krook operator, 
ordering the Boltzmann equation in $\epsilon^1$ and $\epsilon^2$ provides 
\begin{eqnarray}
\label{eq:f1f2}
f_e^{(1)}&=&-{1\over \nu_e}v_x{\partial f_{{\rm m},e}\over \partial x},\\
f_e^{(2)}&=&-{1\over \nu_e}\left({\partial f_{{\rm m},e}\over \partial t} + 
v_x{\partial f_e^{(1)}\over \partial x}\right).
\end{eqnarray}
It shows that the system can be treated as steady state in the first order of 
$\epsilon$ but the secular variation cannot be negligible in the 
second order of $\epsilon$.
The time derivative of the Maxwell-Boltzmann distribution function
can be described as 
\begin{equation}
\label{eq:timder}
{\partial f_{{\rm m},e}\over \partial t}=
-{1\over 2}\left(5-{2v^2\over v_{{\rm th},e}^2}\right)
{1\over T}{\partial T\over \partial t}f_{{\rm m},e} 
%=-\frac{\nu_e \epsilon^2\delta_T}{2}
=-\frac{\epsilon\delta_T}{2}\frac{v_{\rm{th},e}}{L}
\left(5-{2v^2\over v_{{\rm th},e}^2}\right)f_{{\rm m},e}.
\end{equation}
%where the collision frequency $\nu_e$ is treated as constant.

% In the following, we obtain higher order solutions for the distribution 
%function.
%In the first order,
The first- and the second-order solutions for the electron
distribution function are thus obtained as follows:
\begin{eqnarray}
\label{eq:f1}
f_e^{(1)}&=&
%-\frac{\delta_T}{2\nu_e L}v_x
%\left(5 - {m_e v^2\over k_{\rm B} T}\right)f_{{\rm m},e} =
-\epsilon\delta_T
\frac{v_x}{v_{{\rm th},e}}\left(5 - {m_e v^2\over k_{\rm B} T}\right)
f_{{\rm m},e}, \\
\label{eq:f2}
f_e^{(2)}&=&\frac{\epsilon^2 \delta_T}{2}
\left[
\left(5-{2v^2\over v_{{\rm th},e}^2}\right)
%-{v_x^2 v^2\over v_{{\rm th},e}^4}
+\frac{v_x^2}{v_{{\rm th},e}^2}\left(5-{4v^2\over v_{{\rm th},e}^2}\right)
+{v_x^2\over 2 v_{{\rm th},e}^2}
\left(5-{2v^2\over v_{{\rm th},e}^2}\right)^2
\right]f_{{\rm m},e},
\end{eqnarray}
%where we have approximated $\Delta \ln T(x)$ by 
where we have neglected the second derivative of $T(x)$ with respect to $x$
in derivation of the second-order solution.

These equations show that the anisotropy 
in velocity distribution functions are 
induced by the existence of the temperature gradient.
The first-order solution is an odd function of the velocity.
On the other hand, the second-order solution is even in velocity. 
It is straight forward to make sure that the zero background 
electric field assumption is consistent with the zero electric 
current condition under the pressure equilibrium plasma condition,
by checking $\left<f_e \vec{v}\right>=\vec{0}$, 
where $\left< \ \right>$ denotes the average over the velocity. 

By substituting $m_e$ to $m_i$ and $\nu_e$ to $\nu_i=v_i/\lambda_i$
in equations (\ref{eq:f1}) and (\ref{eq:f2}), the first and the second order
distribution functions $f_i^{(1)}$ and $f_i^{(2)}$ for ion are obtained.

\section{Dispersion relation}

In the short-wavelength limit, the collisionless Boltzmann
equation is a good representative for the evolution of 
the perturbed distribution
functions, and the plain wave $\exp[i(\vec{k}\cdot\vec{r}-\omega t)]$ 
is a good representative for the perturbed quantities 
(e.g., electron distribution function
as $\delta f_e=f_{\vec{k},e}\,\exp[i(\vec{k}\cdot\vec{r}-\omega t)]$,
electric field as $\vec{E}=\vec{E}_{\vec{k}}\,
\exp[i(\vec{k}\cdot\vec{r}-\omega t)]$, and magnetic field
as $\vec{B}=\vec{B}_{\vec{k}}\,\exp[i(\vec{k}\cdot\vec{r}-\omega t)]$).
From the linearized Boltzmann equation in terms of the perturbed 
quantities of both electron and ion, the equations 
for the perturbed distribution functions of electron 
and ion  are obtained as
\begin{eqnarray}
\label{eq:lineq}
-i(\omega -\vec{k}\cdot\vec{v}) f_{\vec{k},e}
&=&+{e\over m_e}\left(\vec{E_{\vec{k}}}+\vec{v}\times{\vec{B_{\vec{k}}}\over
c}\right)\cdot \left({\partial f_e\over\partial \vec{v}}\right) \\
-i(\omega -\vec{k}\cdot\vec{v}) f_{\vec{k},i}
&=&-{e\over m_i}\left(\vec{E_{\vec{k}}}+\vec{v}\times{\vec{B_{\vec{k}}}\over
c}\right)\cdot\left({\partial f_i\over\partial \vec{v}}\right).
\end{eqnarray}

As shown below, 
the low-frequency condition requires that the charge density be small,
implying that the electric
current has to be almost transverse for purely electronic 
oscillations. Therefore, only two modes are expected.
Without loss of generality, the wave vector $\vec{k}$ 
can be taken in the $x$-$y$ plane as $k_x=k\cos\theta$, $k_y=k\sin\theta$,
and $k_z=0$.
%$\vec{k}=(k\cos\theta, k\sin\theta, 0)$. 
We expect one mode with the magnetic field along 
the $z$-direction (mode (1): $B_x=B_y=0$ and $B_z\neq 0$), and
the other one with the magnetic field in the $x$-$y$ plane (mode (2):
$B_x,B_y \neq 0$ and $B_z=0$).
We thus denote the component of a vector along the direction of the wave vector
by a $k$-subscript
and the component along the perpendicular direction 
of the wave vector in the $x$-$y$ plain by a $\perp$-subscript.
The $x$-component of velocity, for example, can be expressed 
as $v_x=v_k\cos\theta-v_{\perp}\sin\theta$ in terms of $v_k$ and
$v_{\perp}$.
 
The Faraday's law in Maxwell's equations leads to $E_z=0$ 
(i.e., $\vec E_{\vec k}=(E_k,E_{\perp},0)$) and
$k E_{\perp} = (\omega/c) B_z$ for mode (1).
For mode (2), 
${\rm div}\vec{B}=0$ yields $B_k=0$ (i.e., $\vec B_{\vec k}=(0,B_{\perp},0)$),
and the Faraday's law and $B_z=0$ lead to
$E_{\perp}=0$ and $k E_z=-(\omega/c)B_{\perp}$.
Then we obtain  
\begin{eqnarray}
\label{eq:mod1e}
(\vec{k}\cdot\vec{v}-\omega)f_{\vec{k},e}&=&
-{ie\over m_e}\left[\left(E_k+{v_{\perp}\over c} B_z\right)
{\partial f_e\over \partial v_k} 
+\left(E_{\perp}-{v_k\over c}B_z\right) 
{\partial f_e\over \partial v_{\perp}}\right],\\
\label{eq:mod1i}
(\vec{k}\cdot\vec{v}-\omega)f_{\vec{k},i}&=&+{ie\over
m_i}\left[\left(E_k+{v_{\perp}\over c} B_z\right)
{\partial f_i\over \partial v_k} 
+\left(E_{\perp}-{v_k\over c}B_z\right) 
{\partial f_i\over \partial v_{\perp}}\right]
\end{eqnarray}
for mode (1), and 
\begin{eqnarray}
\label{eq:mod2e}
(\vec{k}\cdot\vec{v}-\omega)f_{\vec{k},e}&=&
-{ie\over m_e}\left[\left(E_k-{v_z\over c} B_{\perp}\right)
{\partial f_e\over \partial v_k} 
+\left(E_z+{v_k\over c}B_{\perp}\right)
{\partial f_e\over \partial v_z}\right],\\
\label{eq:mod2i}
(\vec{k}\cdot\vec{v}-\omega)f_{\vec{k},i}&=&
+{ie\over m_i}\left[\left(E_k-{v_z\over c} B_{\perp}\right)
{\partial f_i\over \partial v_k} 
+\left(E_z+{v_k\over c}B_{\perp}\right)
{\partial f_i\over \partial v_z}\right]
\end{eqnarray}
for mode (2).

To first order in $\omega/(k v_{{\rm th},e})$ and $\omega/(k v_{{\rm th},i})$,
$(\vec{k}\cdot\vec{v}-\omega)^{-1}=(1/k) P(1/v_k)+(\omega/k^2) P(1/v_k^2)
+i(\pi/k)\delta(v_k)-i(\pi/k)(\omega/k) (d\delta(v_k)/ d v_k)$, where 
$P$ denotes the principal value and the 
signs in front of the delta functions 
reflect the causality condition.
Then, with the help of the Faraday's law,  
equations (\ref{eq:mod1e}) and (\ref{eq:mod1i}) yield 
\begin{eqnarray}
\label{eq:fkmod1}
f_{\vec{k},e}=-{ie\over m_e}\left\{
\left[{1\over k}P{1\over v_k}+{\omega\over k^2} P{1\over v_k^2}
+{i\pi\over k}\delta(v_k)-{i\pi\omega\over k^2} 
{d \delta(v_k)\over d v_k}\right]\times \right.\nonumber\\
\left(E_k+{v_{\perp}\over c}B_z\right)
\left(-{2v_k\over v_{{\rm th},e}^2}+\frac{\epsilon\delta_T}{2v_{{\rm th},e}}
\left[
\cos\theta\left(5-{2 v^2\over v_{{\rm th},e}^2}\right)
- {2 v_k\over v_{{\rm th},e}^2}
\left(v_k\cos\theta-v_{\perp}\sin\theta\right)
\left(7-{2 v^2\over v_{{\rm th},e}^2}\right) 
\right]
\right) \nonumber\\
\left.
-{B_z\over ck}
\left(
-{2 v_{\perp}\over v_{{\rm th},e}^2} +\frac{\epsilon\delta_T}{2v_{{\rm th},e}}
\left[-\sin\theta\left(5-{2 v^2\over v_{{\rm th},e}^2}\right)
-{2 v_{\perp}\over v_{{\rm th},e}^2}
\left(v_k\cos\theta - v_{\perp} \sin\theta \right)
\left(7-{2 v^2\over v_{{\rm th},e}^2}\right) 
\right]\right)\right\}f_{{\rm m},e},\\
f_{\vec{k},i}=
+{ie\over m_e}\left\{
\left[{1\over k}P{1\over v_k}+{\omega\over k^2} P{1\over v_k^2}
+{i\pi\over k}\delta(v_k)-{i\pi\omega\over k^2} 
{d \delta(v_k)\over d v_k}\right]\times \right.\nonumber\\
\left(E_k+{v_{\perp}\over c}B_z\right)
\left(-{2v_k\over v_{{\rm th},i}^2}+\frac{\epsilon\delta_T}{2v_{{\rm th},i}}
\left[
\cos\theta\left(5-{2 v^2\over v_{{\rm th},i}^2}\right)
- {2 v_k\over v_{{\rm th},i}^2}
\left(v_k\cos\theta-v_{\perp}\sin\theta\right)
\left(7-{2 v^2\over v_{{\rm th},i}^2}\right) 
\right]
\right) \nonumber\\
\left.
-{B_z\over ck}
\left(
-{2 v_{\perp}\over v_{{\rm th},i}^2} +\frac{\epsilon\delta_T}{2v_{{\rm th},i}}
\left[-\sin\theta\left(5-{2 v^2\over v_{{\rm th},i}^2}\right)
-{2 v_{\perp}\over v_{{\rm th},i}^2}
\left(v_k\cos\theta - v_{\perp} \sin\theta \right)
\left(7-{2 v^2\over v_{{\rm th},i}^2}\right) 
\right]\right)\right\}f_{{\rm m},i}
\end{eqnarray}
for mode (1).
Similarly, from equations (\ref{eq:mod2e}) and (\ref{eq:mod2i}), we obtain
\begin{eqnarray}
\label{eq:fkmod2}
f_{\vec{k},e}
=-{ie\over m_e}
\left\{\left[
{1\over k}P{1\over v_k}+{\omega\over k^2} P{1\over v_k^2}
+{i\pi\over k}\delta(v_k)-{i\pi\omega\over k^2} {d \delta(v_k)\over d v_k}
\right]\right.\times \nonumber\\
\left(-{v_z\over c}B_{\perp}\right)
\left(
-{2v_k\over v_{{\rm th},e}^2}
+\frac{\epsilon\delta_T}{2v_{{\rm th},e}}
\left[\cos\theta\left(5-{2 v^2\over v_{{\rm th},e}^2}\right)
- {2 v_k\over v_{{\rm th},e}^2}
\left(v_k\cos\theta-v_{\perp}\sin\theta\right)
\left(7-{2 v^2\over v_{{\rm th},e}^2}\right) 
\right]\right) \nonumber\\
\left.+{B_{\perp}\over ck}
\left[
-{2 v_z\over v_{{\rm th},e}^2}
-\frac{\epsilon\delta_T}{2v_{{\rm th},e}}
\left(v_k\cos\theta - v_{\perp} \sin\theta \right)
\left(7-{2 v^2\over v_{{\rm th},e}^2}\right)
\right]\right\}f_{{\rm m},e},\\
f_{\vec{k},i}
=+{ie\over m_i}
\left\{\left[
{1\over k}P{1\over v_k}+{\omega\over k^2} P{1\over v_k^2}
+{i\pi\over k}\delta(v_k)-{i\pi\omega\over k^2} {d \delta(v_k)\over d v_k}
\right]\right.\times \nonumber\\
\left(-{v_z\over c}B_{\perp}\right)
\left(
-{2v_k\over v_{{\rm th},i}^2}
+\frac{\epsilon\delta_T}{2v_{{\rm th},i}}
\left[\cos\theta\left(5-{2 v^2\over v_{{\rm th},i}^2}\right)
- {2 v_k\over v_{{\rm th},i}^2}
\left(v_k\cos\theta-v_{\perp}\sin\theta\right)
\left(7-{2 v^2\over v_{{\rm th},i}^2}\right) 
\right]\right) \nonumber\\
\left.+{B_{\perp}\over ck}
\left[
-{2 v_z\over v_{{\rm th},i}^2}
-\frac{\epsilon\delta_T}{2v_{{\rm th},e}}
\left(v_k\cos\theta - v_{\perp}\sin\theta \right)
\left(7-{2 v^2\over v_{{\rm th},i}^2}\right)
\right]\right\}f_{{\rm m},i}
\end{eqnarray}
for mode (2).

The Poisson equation
is written as
\begin{equation} 
\label{eq:poisn}
ikE_k =4\pi e \left(\left<f_{\vec{k},i}\right> - \left<f_{\vec{k},e}\right>
\right).
\end{equation}
Keeping only the first non-vanishing order in $\epsilon$ for mode (1), 
each term on the right-hand side provides
\begin{eqnarray}
\label{eq:fvar}
-4\pi e \left<f_{\vec{k},e}\right>&=& 
-\frac{i}{k}\left[
\frac{2}{\lambda^2_{\rm D}}E_k-\epsilon\delta_T
\frac{\omega^2_{\rm pe}}{cv_{{\rm th},e}}B_z\sin\theta
\right],\\
%- i{8\pi e^2 n_0\over k_{\rm B} T}{1\over k} E_k 
%+ {i\over \nu_e} {4\pi n_0 e^2\over m_e} 
%{\delta_T\over L} \sin\theta \frac{B_z}{c k},\\
%
+4\pi e \left<f_{\vec{k},i}\right>&=& 
-\frac{i}{k}\left[
\frac{2}{\lambda^2_{\rm D}}E_k-\epsilon\delta_T 
\frac{\omega^2_{\rm pi}}{cv_{{\rm th},i}}B_z\sin\theta
\right],
%+4\pi e \left<f_{\vec{k},i}\right>&=& 
%- i{8\pi e^2 n_0\over k_{\rm B} T}{1\over k} E_k 
%+ {i\over \nu_i} {4\pi e^2\over m_i} 
%{\delta_T\over L} \sin\theta {1\over k} n_0 {B_z\over c}.
\end{eqnarray}
where $\omega_{\rm pe}$ and $\omega_{\rm pi}$ are the electron and the
ion plasma frequencies, defined by
$\omega_{\rm pe}\equiv \sqrt{(4\pi n_0 e^2/m_e)}$ and 
$\omega_{\rm pi}\equiv \sqrt{(4\pi n_0 e^2/m_i)}$, respectively;
$\lambda_{\rm D}\equiv \sqrt{{k_{\rm B}T/(4\pi n_0 e^2)}}$
is the Debye length. 
It is trivial why those equations do not contain $E_{\perp}$
since the transverse component is not constrained from the Poisson equation.  
Note that the first term in the right hand side for ion is exactly the same 
as that for electron.
Therefore, the contribution of ion to the charge density is non-negligible. 
On the other hand, the second term for ion is smaller by factor
of $\sqrt{m_e/m_i}$ than that for electron.
This is the essential point to make possible in-phase acoustic  
oscillation between electron and ion.
Then the spatial charge carried by ion acoustic oscillation is
able to be canceled by electron to keep charge neutrality.
In the case of pure electron plasma examined by Ramani \& Laval (1978), 
electrons have to keep charge neutrality by themselves.
Therefore, the amplitude of acoustic oscillation for electron plasma 
has to be almost zero. 
The Poisson equation provides the relation between electric and magnetic 
fields as
\begin{equation}
\label{eq:ekb}
E_k=\frac{\epsilon\delta_T}{4}
{\sin\theta \over 1+(k\lambda_{\rm D})^2}{v_{{\rm th},e}\over c}B_z.
\end{equation}
The term $(k\lambda_{\rm D})^2$ in 
denominator comes from the left-hand side of the Poisson equation and
always very small number since the wavelength of the mode which we are 
interested in is much larger than the Debye length.  
We, therefore,  neglect this term. This corresponds to the so-called Plasma 
approximation (Chen 1974; Tanaka \& Nishikawa 1996) and ensures that
the charge neutrality is kept in high accuracy
even though the longitudinal oscillation exists.  
When the wave vector is parallel to the
temperature gradient ($\theta=0$), 
this equation tells $E_k=0$ and the mode (1) becomes pure transverse.
Compared with the case of pure electron oscillation 
as in Ramani \& Laval (1978), 
the right-hand side of equation (\ref{eq:ekb}) is factor of two smaller, 
which makes mode (1) more unstable as shown in below.
No constraint comes from the Poisson equation for mode (2) since it
is a pure transverse mode. 

We are now in a position to 
derive the dispersion relation.
The Amp\`ere's law is written as 
\begin{equation}
\label{Amp1}
-ikB_z={4\pi e\over c}  \left(\left<v_{\perp}f_{\vec{k},i}\right> - 
\left<v_{\perp} f_{\vec{k},e}\right>\right) +
{\partial E_{\perp}\over 
c\partial t}
\end{equation}
for mode (1), and
\begin{equation}
\label{Amp2}
ikB_{\perp}={4\pi e\over c} \left(\left<v_z f_{\vec{k},i}\right> - 
\left<v_z f_{\vec{k},e}\right>\right) + {\partial E_z\over 
c\partial t}
\end{equation}
for mode(2). 
Under the low-frequency condition, the last terms in the right-hand side
of equations (\ref{Amp1}) and (\ref{Amp2}), 
those are the displacement current, 
are negligibly small. We thus neglect these terms.
Further, the contribution of ion to the current density is order of 
$\sqrt{m_e/m_i}$ smaller than that of electron since the current density 
carried by each particle is proportional to the thermal velocity
of each particle.  Therefore, the current density carried by ion 
introduces only a negligible contribution to the dispersion relation. 
The dispersion relation for the real parts $\omega_{\rm r}$ 
and the imaginary parts $\omega_{\rm i}$
in the leading order of $\epsilon$ is then obtained as follows:
\begin{eqnarray}
\label{eq:disp1a}
\omega_r&=&
\frac{\epsilon\delta_T}{4} k v_{{\rm th},e}\cos\theta,\\
\label{eq:disp1b}
\omega_i&=&
\frac{\epsilon^2\delta_T^2}{4\sqrt{\pi}}
kv_{{\rm th},e}  (2\cos^2\theta -\sin^2\theta) 
-{1\over \sqrt{\pi}}\left( {c\over \omega_{\rm pe}} \right)^2 k^3
v_{{\rm th},e}
\end{eqnarray}
for mode (1), and 
\begin{eqnarray}
\label{eq:disp2a}
\omega_r&=&
\frac{\epsilon\delta_T}{4}v_{{\rm th},e}k {\rm cos}\theta,\\
\label{eq:disp2b}
\omega_i&=&
\frac{\epsilon^2\delta_T^2}{2\sqrt{\pi}}
v_{{\rm th},e} k \cos^2\theta -{1\over\sqrt{\pi}}
\left({c\over \omega_{\rm pe}}\right)^2 
v_{{\rm th},e} k^3,
\end{eqnarray}
for mode (2).

Since the imaginary part $\omega_{\rm i}$ is of order 
$\epsilon^2$, we have checked whether the leading order of the dispersion
relation is changed when the second-order distribution functions 
$f_e^{(2)}$ and $f_i^{(2)}$ are taken into consideration.
We have confirmed that these second-order distribution functions 
only introduce one order higher terms both in real and imaginary parts,
and thus main results shown above are not changed.

\section{Instability and mode characteristics}

The dispersion relations (\ref{eq:disp1a}), (\ref{eq:disp1b}), 
(\ref{eq:disp2a}) and ({\ref{eq:disp2b}) show
that  
the low-frequency mode, for which the phase velocity of the wave is slower 
than ion thermal  velocity, can exist
as long as $\epsilon < \sqrt{m_e/m_i}\sim 0.025$ is satisfied.
This is the first confirmation that the Ramani \& Laval (1978) type 
instability can exist beyond the wall of square root of the mass ratio.   
The dispersion relations are almost the same as those  obtained for
electron plasma by Ramani \& Laval (1978).
However, there is a slight difference in imaginary part of the mode (1).
The difference comes from the difference in the charge neutrality condition
(\ref{eq:ekb}).
As explained in \S 4, in our case, 
the in-phase acoustic oscillation 
between electron and ion is possible. Hence,
the spatial charge carried by ion acoustic oscillation is
able to be canceled by electron to keep charge neutrality.
However, in the case of pure 
electron plasma examined by Ramani \& Laval (1978), 
electrons have to keep charge neutrality by themselves.
Therefore, the amplitude of acoustic oscillation for electron plasma 
has to be almost zero.

The characteristics of the instability are summarized as follows.
For mode (1), 
the imaginary part of the wave frequency 
is positive and the instability sets in,  
when the direction of the wave vector 
is within the double cone spanned by 
$\theta\in[-\theta_{\rm cr},\theta_{\rm cr}]$ and 
$\theta\in[\pi-\theta_{\rm cr},\pi+\theta_{\rm cr}]$,
where $\theta_{\rm cr}\equiv \arccos(1/\sqrt 3)$ ($0\le \theta_{\rm
cr}\le \pi/2$).
%$-\pi/4<\theta<\pi/4$ and $3\pi/4<\theta<5\pi/4$,
% 
For comparison, we re-calculated the dispersion 
relations for pure electron plasma.
In the case of pure electron plasma, 
a factor of $2$ appears in front of
the $\sin^2\theta$ term in the imaginary part
(this result is slightly different from Ramani \& Laval (1978)'s result.).
Therefore, the unstable region in $\vec{k}$-space in the case
of pure electron plasma is somewhat narrower than our case. 
It shows that acoustic oscillation of ion is assisting the instability.
For mode (2), the unstable mode exists for all direction of the wave vector, 
although the growth rate decreases as $\theta$ increase from 0 to $\pi/2$ 
for fixed $k$.
Two modes get identical when $\theta=0$.
For both modes, the growth rate is maximum when $\theta=0$.  
Since the imaginary part of the wave frequency 
is the third order polynomial of $k$ with negative
coefficient for $k^3$, there exists a maximum growth rate 
$\omega_{\rm i,max}$. 
For mode (1), $\omega_{\rm i,max}$ is given as
\begin{equation}
\label{eq:oimaxi1}
\omega_{\rm i,max}\sim {\epsilon^3\delta_T^3\over 12\sqrt{3\pi}}
{v_{{\rm th},e}\over c}\omega_{\rm pe}
%\left|2\cos^2\theta-\sin^2\theta\right|,
\left| 3\cos^2\theta-1 \right|^{3/2},
\end{equation}
when 
\begin{equation}
\label{eq:kimax1}
k=k_{\rm max}={\epsilon\delta_T\over 2\sqrt{3}} 
{\omega_{\rm pe}\over c}
%\left|2\cos^2\theta-\sin^2\theta\right|.
\left|3\cos^2\theta-1\right|^{1/2},
\end{equation}
for  
$\theta\in[-\theta_{\rm cr},\theta_{\rm cr}]$ or 
$\theta\in[\pi-\theta_{\rm cr},\pi+\theta_{\rm cr}]$. 
For mode (2), 
\begin{equation}
\label{eq:oimax2}
\omega_{\rm i,max}\sim {\epsilon^3\delta_T^3\over 3\sqrt{6\pi}}
{v_{{\rm th},e}\over c}\omega_{\rm pe}\left|\cos^3\theta\right| 
\end{equation}
when
\begin{equation}
\label{eq:kimax2}
k=k_{\rm max}={\epsilon\delta_T\over \sqrt{6}} 
{\omega_{\rm pe}\over c}\left|\cos\theta\right|,
\end{equation}
for arbitrary $\theta$.
For the wave vector which satisfies the above conditions, 
the real part $\omega_{\rm r}$ of the wave frequency is
obtained as 
\begin{equation}
\label{eq:ormax1}
\omega_{\rm r}\sim {\epsilon^2\delta_T^2\over 8\sqrt{3}}
{v_{{\rm th},e}\over c}\omega_{\rm pe}
%\left| 2\cos^2\theta-\sin^2\theta \right|
\left| 3\cos^2\theta-1 \right|^{1/2}
\cos\theta
\end{equation}
for mode (1), and
\begin{equation}
\label{eq:ormax2}
\omega_{\rm r}\sim {1\over 4\sqrt{6}}\epsilon^2\delta_T^2 {v_{{\rm th},e}
\over c}\omega_{\rm pe}\cos^2\theta
\end{equation}
for mode (2).

The electric field strength can be related to the magnetic field
strength.
For mode (1),  
\begin{equation}
\label{eq:Ekstren}
|E_k| \sim {\epsilon\delta_T\over 4} {v_{{\rm th},e}\over c}
|\vec B_{\vec k}|
|\sin\theta|.
\end{equation}
When $k=k_{\rm max}$, the electric field perpendicular to the $\vec
k$-direction is related to the magnetic field as
\begin{equation}
\label{eq:Epstren}
|E_{\perp}|\sim {\epsilon\delta_T\over 4}{v_{{\rm th},e}\over c}
|\vec B_{\vec k}|
|\cos\theta|.
\end{equation}
For mode (2), 
\begin{equation}
\label{eq:Ezstren}
|E_z|\sim {\epsilon\delta_T\over 4}{v_{{\rm th},e}\over c}
|\vec B_{\vec k}|
|\cos\theta|,
\end{equation}
when $k=k_{\rm max}$.

The electric field strength is then order of $\epsilon$ smaller than the
magnetic field strength. 
This nature combined with the low-frequency nature of the mode 
shows that the mode has  similarity  
with the magneto-hydrodynamical mode. 
However,  the magneto-hydrodynamical treatment cannot identify the mode and the instability.
For example, the instability is microscopical, in which case
the resonance of particles with waves is essential.

The possible reduction factor of the mean free path, in other words 
the possible lowest value of $\epsilon$, is set by the application
limits and assumptions summarized in \S 2. 
In the following discussion, we treat the fastest growing mode,
that is $\theta=0$ case.
First, the condition (\ref{eq:cond4a}) provides
\begin{equation}
\label{eq:maincon}
\epsilon > 4 \left({1\over L}{c\over \omega_{\rm pe}}\right)^{1/2}
\left({\delta_T\over 1.0}\right)^{-3/2}.
\end{equation}
Second, the age condition (\ref{eq:cond5}) provides
\begin{equation}
\label{eq:agecon}
\epsilon > 2 \left({H_0\over v_{{\rm th},e}}{c\over\omega_{\rm pe}}
\right)^{1/3}\left({\delta_T\over 1.0}\right)^{-1}.
\end{equation}
Third, the collisionless condition is satisfied when
$2\pi/k_{\rm max}< \epsilon \times L$, which yields
\begin{equation}
\label{eq:cless}
\epsilon > 
4 \left({1\over L} {c\over\omega_{\rm pe}}\right)^{1/2}
\left({\delta_T\over 1.0}\right)^{-1/2}.
\end{equation}
%The final condition is almost equivalent to the condition of 
%Eq.(\ref{eq:maincon}) and 
%is always satisfied when the condition of Eq.(\ref{eq:maincon}) is satisfied. 
Finally, the condition that the life time of the interface 
given by equation (\ref{eq:diftim}) 
must be  longer than the age of the cool region, $t_{\rm age}(\rm cool)$, 
provides 
the upper limit on $\epsilon$ as  
\begin{equation}
\label{eq:liftim}
\epsilon < 10^{-5} \left({t_{\rm age}(\rm cool)\over 10^{10}
{\rm yr}}\right)^{-1} 
\left({L\over 4{\rm kpc}}\right) \left({k_{\rm B}T_e\over 
4{\rm keV}}\right)^{-0.5}.
\end{equation}

\section{Application to the cluster of galaxies}

\subsection{A jam bun model}

For the several nearby  
clusters of galaxies,  the existence of at least the 
two different temperature gas in the central region
is spectro-scopically confirmed (Ikebe 1995).
If the cool gas found in the cluster central region is 
a single block of the low temperature gas located at the bottom of
the cluster gravitational potential well like a jam in a jam bun,
the interface between the cool and hot gas can be well described
by the pressure equilibrium condition and 
the results obtained in this paper can be applied.
One of the well studied such clusters is the Centaurus cluster 
(Ikebe et al. 1999).
In the following discussion, the cool gas found in the Centaurus cluster is 
assumed to be well described by the jam bun model.
The temperature of 
the cool region is about $k_{\rm B}T_1\sim1$keV and that of 
the hot region is about $k_{\rm B}T_2\sim$4keV in the Centaurus cluster 
(Ikebe et al. 1999).
The factor $\delta_T$ is obtained as 
$\delta_T\sim (T_2-T_1)/(T_2+T_1)/2 \sim 1.2$.
The size of the region, where the cool component exists, 
is about the size of the X-ray halo around the elliptical galaxies, 
say 50-100$h^{-1}$kpc (Forman, Junes, \& Tucker 1985; Matsushita et
al. 1998; Ikebe et al. 1999). 
Using the best fit values for the electron density profile
for the hot phase gas obtained by ASCA results (Ikebe et al. 1999), 
the electron number density of the hot phase gas at around 100kpc
can be estimated as $2.5\times 10^{-3}{\rm cm^{-3}}$.
Then the electron mean free path $\lambda_{\rm C}$ 
due to Coulomb collisions is 
about 2kpc. Hence,  
the condition (\ref{eq:coulomb}) implies $L>2$kpc, at least 
when the low-temperature region was formed. 
The age condition (\ref{eq:agecon}) leads to
\begin{equation}
\label{eq:age}
\epsilon > 4 \times 10^{-7} h^{1/3}\left({k_{\rm B}T_e\over 4{\rm keV}}
\right)^{-1/6}
\left({n_0\over 2.5\times 10^{-3}{\rm cm^{-3}}}\right)^{-1/6}
\left({\delta_T\over 1.2}\right)^{-1}.
\end{equation}
%
%The condition of Eq.(\ref{eq:maincon}) provides a 
%lower limit on $\epsilon$ as
The collisionless condition (\ref{eq:cless}) also provides a lower limit
on $\epsilon$ as
%\begin{equation}
%\label{eq:cen1}
%\epsilon > 7.7 \times 10^{-8} \left({n_0\over 2.5\times 10^{-3}{\rm cm^{-3}}}
%\right)^{-1/4}
%\left({L\over 4{\rm kpc}}\right)^{-1/2}\left({\delta_T\over 1.2}
%\right)^{-1.5}.
%\end{equation}
\begin{equation}
\label{eq:cen1}
\epsilon > 10^{-7} \left({n_0\over 2.5\times 10^{-3}{\rm cm^{-3}}}
\right)^{-1/4}
\left({L\over 4{\rm kpc}}\right)^{-1/2}\left({\delta_T\over 1.2}
\right)^{-1/2}.
\end{equation}
Therefore, the age condition yields the lower limit on $\epsilon$ when 
$L> 0.25$kpc and is compatible with the life time 
condition (\ref{eq:liftim}).
On the other hand, 
the condition (\ref{eq:cen1}) provides the
lower limit on $\epsilon$ when $L<0.25$kpc.
In this case, 
for this condition to be compatible with 
the life time condition (\ref{eq:liftim}),
the lower limit of the size of the interface region is set as
%\begin{equation}
%\label{eq:lcon}
%L> 0.7 \left({t_{\rm age}(\rm cool)\over 10^{10}{\rm yr}}\right)^2 
%\left({n_0\over 2.5\times 10^{-3}{\rm cm^{-3}}}\right)^{-1/2}\left({k_{\rm B}
%T_e\over 4{\rm keV}}\right)
%\left({\delta_T\over 1.2}\right)^{-3}{\rm pc}.
%\end{equation}
\begin{equation}
\label{eq:lcon}
L> 0.2 \left({t_{\rm age}(\rm cool)\over 10^{10}{\rm yr}}\right)^{2/3} 
\left({n_0\over 2.5\times 10^{-3}{\rm cm^{-3}}}\right)^{-1/6}\left({k_{\rm B}
T_e\over 4{\rm keV}}\right)^{1/3}
\left({\delta_T\over 1.2}\right)^{-1/3}{\rm kpc}.
\end{equation}
Compared with 
the cluster pressure scale height,
the obtained small values of the lower limit on $L$ both from the age 
condition and the condition (\ref{eq:lcon})
%, compared with the cluster pressure scale height, 
guarantee 
the application of the pressure equilibrium condition. 
Therefore, we conclude that the instability found in this paper can 
be a possible mechanism to inhibit the  heat conduction between
cool and hot phases in the cluster central region, if 
the cool gas is 
a single block of the low temperature gas located at the bottom of the
cluster gravitational potential well.
%It may be worth to note that the thickness of the interface region $L$ 
%is able to become 
%smaller than the Coulomb collision electron mean free path as $\epsilon$ gets 
%smaller by the instability.

\subsection{The raisin bread model}

If the cool gas is contained in small blobs and the blobs 
distribute in the entire cluster central region like raisins
in a raisin bread, 
the obtained results cannot be simply applied.
The cooling flow model, 
which is 
one of the widely-accepted model to describe 
the cluster central region (Fabian 1994), 
assumes that the ICM in 
the cluster central region is in multi-phase and 
is composed of many small cool clumps (raisins) which have 
different temperatures from clump to clump. 
Even in the two temperature phase model (Ikebe et al. 1999), 
the raisin bread model is thought to be 
more plausible to describe the situation in the cluster central region.
It is well known that these cool clumps, which is in pressure 
equilibrium with surrounding hot medium in the cluster central region,
oscillates owing to  buoyancy force 
(Balbus 1986; Malagoli, Rosner and Bodo 1987;
Balbus and Soker 1989) and that they are fragmented (Nulsen 1986; Balbus
1986)  and merge into the surrounding hot medium within a few 
buoyancy oscillation period, 
that is $\sim {\rm a \;\;few}\times10^8$yr (Hattori and Habe 1990; 
Malagoli, Rosner and Fryxell 1990; Reale et al. 1991; 
Yoshida, Hattori and Habe 1991).
It has been shown that the magnetic field frozen in the ICM 
can prevent the buoyancy oscillation of the cool clumps
(Nulsen 1986; Loewenstein 1989; Balbus 1991; Hattori, Yoshida and Habe
1995). 
Therefore, the raisin bread model requires 
the existence of the magnetic field in the plasma.
Nulsen (1986) pointed out  
that if the Alfv\'en velocity $v_{\rm A}$ of the hot medium
is larger than the terminal velocity $v_t$ of the cool clumps,
that is, if
\begin{equation}
\label{eq:Nulsen}
v_{\rm A}>v_t,
\end{equation}
the relative motion of cool clumps against the
surrounding medium is stopped and the fragmentation of the
clumps are prevented.
The terminal velocity of a cool  clump
with a relative density excess of $\delta\rho$ and
a size of $r$ is given by $v_t=\sqrt{(\delta \rho/\rho)(r/R)}$,
where $R$ is the pressure scale height of the cluster.
It was confirmed by the numerical simulation (Hattori,
Yoshida and Habe 1995) that this condition works well even
for the cool clumps with $\delta\rho/\rho>1$. 
It is important to note that even if 
the magnetic field line is parallel to the 
gravitational force, the buoyancy oscillation can be 
prevented by the magnetic tention which is 
first pointed out by Loewenstein (1989).

It can be shown that the gas pressure equilibrium condition
is a good approximation even for the cool clumps 
in the raisin bread model. 
Suppose that a cool gas clump is in pressure equilibrium  
with surrounding medium which is immersed in the 
background magnetic field. 
The cool gas clump falls ahead of the surrounding medium 
owing to its excess weight. 
As a result, the magnetic field configuration of the 
surrounding medium is locally disturbed.
The infall of the cool clump is stopped, 
when the force produced by the 
disturbed magnetic field acting on the cool clump
balance with the buoyancy force.
This condition is written as 
\begin{equation}
\label{eq:balance}
{P_m\delta_B\over r} \sim \delta_{\rho} {P_g\over R},
\end{equation}
where $P_m$ and $P_g$ are magnetic and gas pressures of the 
hot surrounding medium, respectively; 
$\delta_B=\delta B/B$ and $\delta_{\rho}=\delta\rho/\rho$ 
are deviations of the magnetic field strength and the gas density 
in the clump, respectively, relative to  
those of surrounding medium when the 
force balance is realized.
This condition can be rewritten  using the 
plasma $\beta$ ($\beta\equiv P_g/P_m$) as 
\begin{equation}
\label{eq:beta}
\delta_B\sim \beta\delta_{\rho} {r\over R}.
\end{equation} 
From equation (\ref{eq:beta}),
the condition (\ref{eq:Nulsen}) 
can be interpreted as $1>\delta_B$, and
the condition for the possible cool clumps in the fixed-$\beta$ plasma
is obtained as 
\begin{equation}
\label{eq:Nulsen2}
{1\over \beta}>\delta_{\rho} {r\over R}.
\end{equation}
In the equilibrium state, the total pressure in the 
clump must be the same as the total pressure of the 
surrounding hot medium. 
This condition provides
\begin{equation}
\label{eq:preeq}
|\delta P_g| \sim |\delta_B P_m|,
\end{equation}
in the clump.
Therefore, the deviation of the gas pressure relative to 
the surrounding gas pressure can be estimated as 
\begin{equation}
\label{eq:devPg}
{|\delta P_g|\over P_g} \sim \delta_{\rho} {r\over R}.
\end{equation} 
From equations (\ref{eq:Nulsen2}) and (\ref{eq:devPg}), 
we obtain the constraint on the relative deviation of the gas pressure: 
\begin{equation}
\label{eq:Pgconst}
{1\over \beta}>{|\delta P_g|\over P_g}.
\end{equation}
Since the hot gas in the cluster is high-$\beta$ plasma, say
$\beta\sim 10-100$ (Kim et al. 1989; Makino 1997), 
the relative deviation of the gas pressure 
%from the gas pressure of the surrounding medium 
is at most 10\% and very small. 

The arguments presented above
ensures that the model of the interface adopted in this 
paper is applicable to the case when the raisin bread model
is the good representative for the cluster central region.
Therefore, the results obtained in this paper 
can provide some hint for reduction of the 
heat conductivity in cooling flow. 
However, the further analysis including the background 
magnetic field is required to apply the raisin 
bread model since the background magnetic field introduces
another resonance related to cyclotron frequencies of electron
and ion.

\section{Discussion}

We have shown that there exists low-frequency growing modes
driven by a global temperature gradient in electron and ion 
plasmas. 
The instability provides a hint for  reduction mechanism of the thermal 
conductivity to explain the
observational evidence of co-existence of the multi-phase gas
in clusters of galaxies. 
However, to answer whether the instability really reduces the 
heat conductivity down to the required level, 
non-linear saturation level of the instability 
and the effect of the back ground magnetic field must be examined. 
Although Levinson \& Eichler (1992) has studied the effect of the 
magnetic field on the Ramani and Laval type instability for pure
electron plasma, there is no study which includes ion with
back ground magnetic field.

The evolutionary nature of the modes in the interface region 
can be briefly described as follows. 
When the low-temperature gas is injected into the intracluster medium, 
any wave is not excited in the interface region yet. Hence, $\epsilon$ is 
defined by the mean free path due to Coulomb collisions 
and is order of 1 to 0.1,
as shown in \S 1. 
At this initial time period, the Ramani \& Laval (1978) type instability 
excites the waves only for electron plasma, since 
the phase velocity of excited waves is about the 
electron thermal velocity and thus much faster than the ion thermal velocity.
As waves grow, the mean free paths of electron and ion become shorter
and $\epsilon$ gets smaller. 
When $\epsilon$ goes down to $\sqrt{m_e/m_i}$, 
ions start to respond the waves, and the modes 
switch to those described in this paper. 
Since the growth rate of the instability is also proportional to
$\epsilon^3$ and very large at the initial stage, 
the condition (\ref{eq:maincon}) is satisfied. 
We thus do not have to worry about the disappearance of the interface 
due to the thermal conduction before the waves grow. 
The waves must be scattered back at the edge of the interface and 
must be confined in the interface region,
because the waves driven by the instability cannot propagate
in the uniform temperature region.

Since the instability creates and grows the magnetic field 
without any seed field, it provides a new possibility  
for the origin of the cosmic magnetic field. 
The Biermann battery effect (Biermann 1950) is the
mechanism which has been 
considered as one of the central mechanism to 
create the seed magnetic field in our universe (Kulsrud et al. 1997). 
The Biermann battery effect requires that
the direction of the density gradient of the gas is not parallel to the 
direction of the temperature gradient.
However, the instability found in this paper requires 
only the existence of the temperature gradient.  
In the astrophysical situation, the situation that two different 
temperature plasmas are
contacting keeping pressure equilibrium can be commonly found compared 
with the situation that the Biermann battery effect may act. 
Therefore it may be interesting to examine how the instability found 
in this paper plays a role on the origin of the cosmic magnetic field.

Finally, it is interesting to note 
that our results suggest that the realization of the global thermal 
equilibrium is postponed 
by the local instability which is induced for quicker 
realization of local thermal equilibrium state in plasmas. 
This result is of course not trivial.  Once the global thermal equilibrium is 
achieved, the local thermal equilibrium is realized simultaneously.
Therefore, letting electrons be free to conduct thermal energy without 
any instability is the fastest way to achieve thermal equilibrium.  
However, the plasma prefers restoring the local deviation of 
the velocity distribution function from the Maxwell-Boltzmann
and excites waves by the instability. As a result,
the realization of the local thermal equilibrium is also 
postponed.  The plasma be hurried into an error. 

\acknowledgments{
Authors would like to thank Masahiro Hoshino, Rudolf Treumann, 
Yousuke Itoh, Yutaka Fujita, Masahide Iijima for many instructive and
 fruitful discussion
on this topics.}
%%%%%%%%%%%%%%%%%%%%%%%%%%%%%%%%%%%%%%%%%%%%%%%%%%%%%%%%%%%%%%%%%%%   

\clearpage
\begin{figure}
\vspace{110mm}
\special{hattori_umetsu_fig1.ps hoffset=0 voffset=350 hscale=0.6 vscale=0.6 rotation=-90}
%%The number density distribution $n(x)$ is also 
%% shown dashed. The doted line shows the distribution of pressure, which
%% is constant along the $x$-direction.}
%\end{figure}
\figcaption{A sketch of one-dimensional distribution of temperature $T(x)$
 in the interface region. $T_1$ and $T_2$ are the temperatures in the
 cool and hot regions, respectively.
Also shown are the number density distribution $n(x)$ (dashed line)
and the distribution of pressure (dotted line), which is constant along
 the $x$-direction. } 
\end{figure}

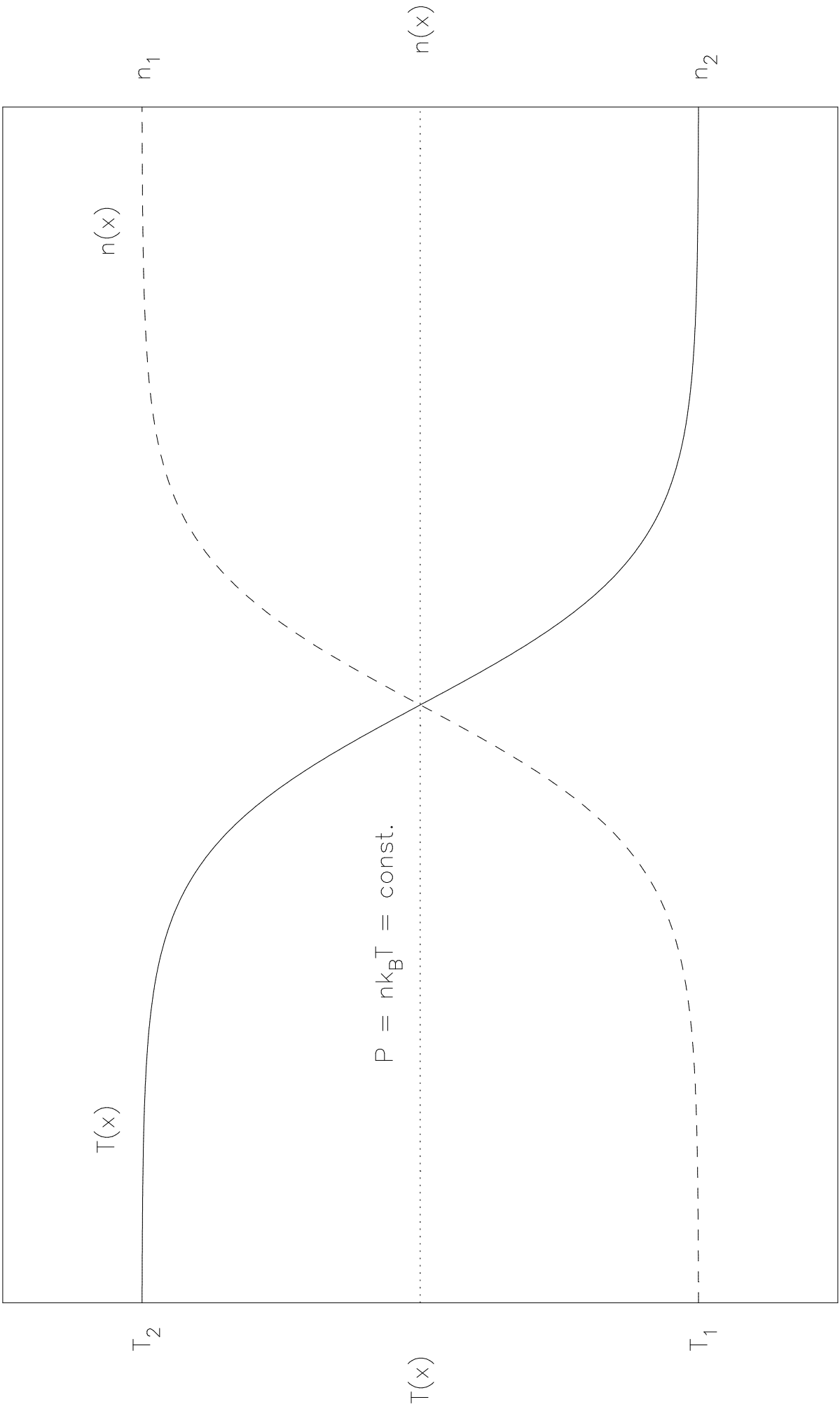
\end{document}